\documentclass[aps,prb,twocolumn,groupedaddress,amsmath,amssymb]{revtex4-1}

\usepackage{graphicx}
\usepackage{subfigure}
\usepackage{epstopdf}
\usepackage{hyperref}
\usepackage{bm}

\begin{document}

\title{Tilting of the magnetic field in Majorana nanowires: critical angle and zero-energy differential conductance}

\author{Stefan Rex}
\author{Asle Sudb{\o}}
\affiliation{Department of Physics, Norwegian University of Science and Technology, N-7491 Trondheim, Norway}

\date{\today}

\begin{abstract}
Semiconductor nanowires with strong spin-orbit coupling and proximity-induced $s$-wave superconductivity in an external magnetic field have been the most promising settings for approaches towards experimental evidence of topological Majorana zero-modes. We investigate the effect of tilting the magnetic field relative to the spin-orbit coupling direction in a simple continuum model and provide an analytical derivation of the critical angle, at which the topological states disappear. We also obtain the differential conductance characteristic of a junction with a normal wire for different tilting angles and propose a qualitative change of the dependence of the zero-energy differential conductance on the tunnel barrier strength at the critical angle as a new criterion for establishing the topological nature of the observed signal.
\end{abstract}

\maketitle

\section{Introduction}
Many decades after the prediction of Majorana fermions\cite{Majorana}, with no direct and unequivocal experimental evidence for their existence, the possibility of finding emergent Majorana modes of a topological nature in condensed matter systems has evoked considerable interest in a number of systems\cite{ReadGreen2000,Kitaev2001,DasSarma2006, Tsutsumi2008, Gurarie2005, FuKane2008, Wilceck2009, Linder2010, Alicea2012, Tanaka2012, Nakosai2013, Asano2013}, partly because of their expected non-Abelian braiding statistics\cite{ReadGreen2000, Ivanov2001, Kitaev2003}. Among the proposed systems, semiconductor nanowires\cite{LutchynSauDasSarma2010} with strong spin-orbit coupling (SOC) and induced $s$-wave superconductivity in an external magnetic field (Majorana nanowires) have become the most prominent setting. Here, suspected signatures of Majorana zero-modes have already been measured\cite{DelftExp, Rokhinson2012, Das2012, Finck2013, Deng2012, Churchill2013, Lee2013}. However, the experimental findings do not match the predictions precisely, and some predictions therefore have been made for more realistic nanowire models\cite{LinSauDasSarma2012,LimLopezSerra2012,LimLopezSerra2013,OscaSerra2013,Prada2012}. This includes, for instance, finite temperature, finite-size effects and the three-dimensional wire geometry. Still, further distinguishing criteria for the existence of the topological states in experiment are desirable.
\par In the present work, we go back to a simple and analytically accessible one-dimensional continuum model. We focus on the possibility of driving the topological phase transition by changing the direction of the magnetic field relative to the SOC direction, while the standard choice is taking them orthogonal. It is immediately clear that the Majorana zero-modes cannot exist for arbitrary field directions. Some experiments have included a rotation of the external magnetic field, but there has been limited quantitative analysis \cite{OscaRuizSerra2014, LinSauDasSarma2012} of the precise impact of the field direction on the Majorana zero-modes and the measured quantity, namely the differential conductance in a junction of the Majorana nanowire with a normal lead.
\par In this paper, we carry out a detailed analysis of the effect of rotating the magnetic field, with particular emphasis on identifying features of the differential conductance directly connected to the the topological character of the zero-energy modes. In Sec.~\ref{SecHamiltonian}, we formulate the Hamiltonian of the system. In Sec.~\ref{SecCriticalAngle}, we present a way to analytically derive the allowed field directions in terms of a critical angle, for which the system remains in the topological phase. Our analytical results confirm the numerically inspired results of Ref.~\onlinecite{OscaRuizSerra2014}. In Sec.~\ref{SecConductance}, we compute the differential conductance characteristics of a normal-Majorana nanowire junction for various angles of the Zeeman field relative to the spin-orbit coupling direction. In particular, we concentrate on the zero-energy differential conductance and propose one further criterion for testing the topological origin of the observed peak by varying the tunnel barrier strength while tilting the field across the critical angle. The main result is that below some critical tilting angle away from the direction where the Zeeman-field and the SOC are orthogonal, the value of the zero-energy peak is quantized in units of $2 \frac{e^2}{h}$, where $e$ is the electron charge and $h$ is Planck's constant, independent of the tunnel barrier of the junction, the value being protected by topology. Beyond a certain angle, this is no longer so, and the value of the zero-energy peak depends on the barrier potential. Conclusions are given in Sec.~\ref{SecConclusion}. 

\section{Model Hamiltonian}\label{SecHamiltonian}
\begin{figure}
\includegraphics[width=0.8\columnwidth]{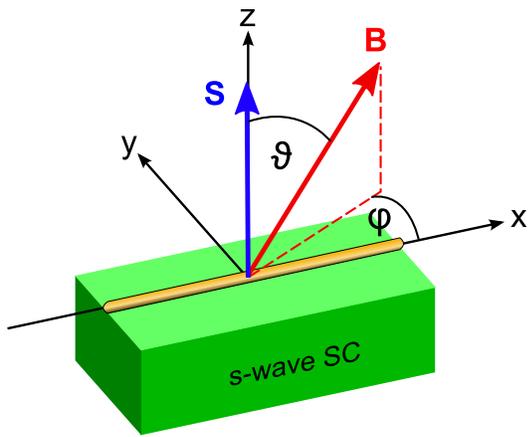}
\caption{\label{FigSketch}Schematic view of the system: The semiconductor nanowire (yellow) is placed on a bulk $s$-wave superconductor and defines the $x$-axis of the coordinate system. The $z$-axis is parallel to the SOC direction (labeled ${\bm S}$ in the figure). The direction of the magnetic field ${\bm B}$ is represented by the two angles $\vartheta$ (tilting relative to the SOC) and $\varphi$ (azimuthal rotation in the $xy$-plane).}
\end{figure}

We consider a one-dimensional semiconductor nanowire with SOC strength $\alpha$ and a proximity-induced $s$-wave superconducting gap $\Delta$. Thermal effects can be taken into account in a simple way by taking into account the temperature dependence of the gap $\Delta$ in the standard way, at least for temperatures not too close to the superconducting transition temperature. 
In this paper, we choose the nanowire to be aligned with the $x$-axis, with the SOC in $z$-direction. We express the external magnetic field ${\bm B}$ in spherical coordinates, with the polar angle $\vartheta$ measured from the $z$-axis and the azimuthal angle $\varphi$ measured from the $x$-axis, and introduce the Zeeman energy $E_\mathrm{Zee} = \frac{1}{2}g\mu_\mathrm{B}B$. A sketch of the system and the chosen coordinates can be found in Fig.~\ref{FigSketch}.
The Bogoliubov-De~Gennes (BdG) Hamiltonian acting on spinors $\psi=(u_\uparrow, u_\downarrow, v_\uparrow, v_\downarrow)^\mathrm{T}$, where $u, v$ refer to the electron and hole part of a quasiparticle and $\uparrow, \downarrow$ to the spin in $z$-direction, respectively, reads\cite{LutchynSauDasSarma2010}
\begin{subequations}\label{EqHamiltonian}
\begin{equation}
   H(k) = \begin{pmatrix}
	          h_\mathrm{n}(k)          & h_\mathrm{sc}(k)          \\
						h_\mathrm{sc}^\dagger(k) & -h_\mathrm{n}^\mathrm{T}(-k)
	        \end{pmatrix}
,\end{equation}
with the normal part
\begin{equation}
   h_\mathrm{n}(k) = \begin{pmatrix}
	                     \xi_k+E_\mathrm{Zee}\cos\vartheta+k\alpha & E_\mathrm{Zee}\sin\vartheta e^{-i\varphi} \\
											 E_\mathrm{Zee}\sin\vartheta e^{i\varphi}  & \xi_k-E_\mathrm{Zee}\cos\vartheta-k\alpha
	                   \end{pmatrix}
\end{equation}
and $s$-wave pairing
\begin{equation}
   h_\mathrm{sc}(k) = h_\mathrm{sc} = \begin{pmatrix}
	                                      0 & \Delta \\ -\Delta & 0
	                                    \end{pmatrix}
,\end{equation}
\end{subequations}
where $\xi_k=(\hbar k)^2/2m-\mu$, $m$ is the effective electron mass, and $\mu$ the chemical potential.

\section{Critical Angle}\label{SecCriticalAngle}
\begin{figure}
\includegraphics[width=\columnwidth]{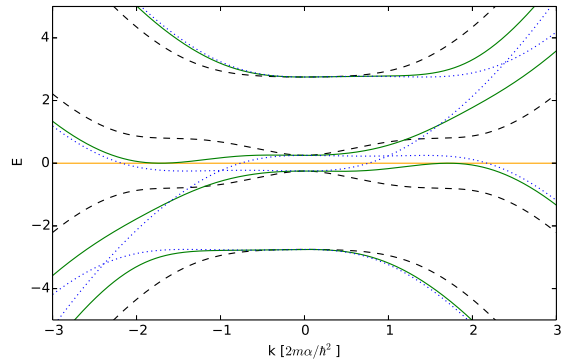}
\caption{\label{FigEnergyLevels} The four eigenenergies of the BdG Hamiltonian Eq.~\eqref{EqHamiltonian} as a function of momentum for $\vartheta=\frac{\pi}{2}$ (black dashed lines), at the critical angle (green solid lines), where the gap closes (here $\vartheta_c\approx0{.}81\pi$), and at $\vartheta=\pi$ (blue dotted lines). The orange line indicates zero energy. Parameters: $m=1, \Delta=1{.}25, E_\mathrm{Zee}=1{.}5, \alpha=\sqrt{1/2}, \mu=0$.}
\end{figure}

It is well-known theoretically that the system harbors Majorana zero-modes in the topological phase, $E_\mathrm{Zee} > \sqrt{\Delta^2+\mu^2}$\cite{Kitaev2001, LutchynSauDasSarma2010, OregRefaelOppen2010}, when ${\bm B}$ is orthogonal to the SOC direction ($\vartheta=\frac{\pi}{2}$). If the field is tilted, on the other hand, the Majorana modes disappear at a critical angle \cite{LinSauDasSarma2012, OscaRuizSerra2014} $\vartheta_c$, where the energy gap closes. Figure \ref{FigEnergyLevels} illustrates the eigenenergies of the BdG Hamiltonian Eq.~\eqref{EqHamiltonian} for parallel and orthogonal field and at $\vartheta=\vartheta_c$. We note that level crossings happen only at $\vartheta=\pi$, thus the gap closes only indirectly at $\vartheta_c$.
The second angle, $\varphi$, only gives a phase factor in the eigenstates and is irrelevant for the eigenenergies and the discussion of topological states. The critical angle was observed to follow a rule equivalent to $\cos\vartheta_c=\Delta/E_\mathrm{Zee}$ in numerical calculations\cite{OscaRuizSerra2014}\footnote{With our choice of coordinates, the rule contains only one instead of two angles, and appears with a cosine instead of a sine.}. In this section, we provide the analytical derivation of this rule.
\par
\begin{figure}
\includegraphics[width=\columnwidth]{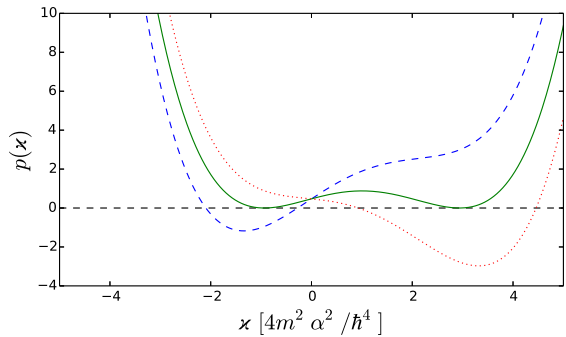}
\caption{\label{FigPolynomial}The characteristic polynomial $p(\varkappa)$ 
of the Hamiltonian at zero energy as a function of $\varkappa=k^2$ for the tilting angles $\vartheta=0{.}74\pi$ (blue dashed line), the critical angle $\vartheta_c\approx0{.}81\pi$ (green solid line), where positive solutions for $\varkappa$ appear first, and $\vartheta=0{.}9\pi$. Parameters: $m=1, \Delta=1{.}25, E_\mathrm{Zee}=1.5, \alpha=\sqrt{1/2}, \mu=0$.}
\end{figure}

 Technically, the task is to find the angle at which the low-energy band first reaches zero energy. The calculation of the eigenenergies is done via the characteristic polynomial, $p_k(E)=\det(H(k)-E)$, which is of order 8 in momentum. For $E=0$, all odd powers of $k$ vanish, leaving a biquartic equation. With the substitution $\varkappa=k^2$, it reads
\begin{eqnarray}\label{EqPolynomial}
p(\varkappa) &=& \left[\left(\frac{\hbar^2}{2m}\varkappa-\mu\right)^2
               -\alpha^2\varkappa+\Delta^2-E_\mathrm{Zee}^2\right]^2 \notag\\
						&& + 4\alpha^2(\Delta^2-E_\mathrm{Zee}^2\cos^2\vartheta)\varkappa.
\end{eqnarray}
As long as the band gap remains open, $p_k(0)$ will be solved only by complex momenta, whereas real solutions appear when ${\bm B}$ is tilted beyond the critical angle. The real solutions of $p_k(0)$ lead to non-negative solutions of $p(\varkappa)$. To derive the critical angle, we will exploit the special form of Eq.~\eqref{EqPolynomial}, being the square of a quadratic polynomial in $\varkappa$, with one additional $\varkappa$-linear term containing the dependence on $\vartheta$. We analyze the quadratic expression first, and find its zeros
\begin{eqnarray}\label{EqSolutionQuadratic}
\varkappa_{1,2} &=& \frac{1}{2}\left(\frac{2m}{\hbar^2}\right)^2\left[\frac{\hbar^2\mu}{m}+\alpha^2\pm\right.\notag\\
&&\left.\sqrt{\left(\frac{\hbar^2\mu}{m}+\alpha^2\right)^2-\left(\frac{\hbar^2}{m}\right)^2(\mu^2+\Delta^2-E_\mathrm{Zee}^2)}\right].
\end{eqnarray}
To allow for topological states at all, $(\mu^2+\Delta^2-E_\mathrm{Zee}^2)$ must necessarily be negative\cite{Kitaev2001, LutchynSauDasSarma2010, OregRefaelOppen2010}. Thus, Eq.~\eqref{EqSolutionQuadratic} always yields two real solutions, where $\varkappa_1>0$ and $\varkappa_2<0$.
In the absence of the linear term, Eq.~\eqref{EqPolynomial} is positive semidefinite and will have precisely the same solutions, just two-fold degenerate each. If, however, the $\varkappa$-linear term is present with positive (negative) coefficient, the point-symmetry of $p(\varkappa)$ is lost and the solutions become non-degenerate, where the positive solution is split in two distinct complex (real) values, cf. Fig~\ref{FigPolynomial}. We conclude from Eq.~\eqref{EqPolynomial} that the system is in the topological phase, when $\Delta^2-E_\mathrm{Zee}^2\cos^2\vartheta>0$. Consequently, the critical angle satisfies
\begin{equation}
\cos\vartheta_c = \pm\frac{\Delta}{E_\mathrm{Zee}}.
\label{Crit_Ang}
\end{equation}
Thus, we have analytically confirmed the numerical results obtained in Ref.~\onlinecite{OscaRuizSerra2014}. As the angle $\vartheta$ is increased through the value $\vartheta_c$, topologically trivial zero-energy states will appear with the momentum $\pm\sqrt{\varkappa_1}$. An alternative, but much more lengthy, derivation of the same result using the discriminant\cite{Rees1922} of the fourth-order polynomial $p(\varkappa)$, is also possible.
\par
\begin{figure}
\includegraphics[width=0.7\columnwidth]{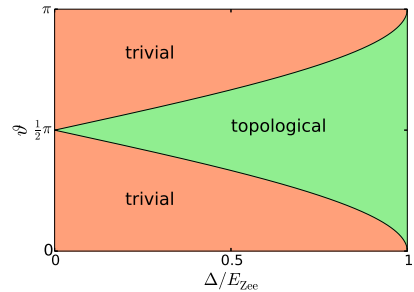}
\caption{\label{FigPhaseDiagram}The angle-resolved topological phase diagram of the Majorana nanowire.}
\end{figure}

 The angle-resolved topological phase diagram is shown in Fig.~\ref{FigPhaseDiagram}. If the Zeeman energy is just slightly larger than the superconducting gap, $\vartheta$ can be varied over a wide range without destroying the Majorana zero-modes, whereas for large Zeeman energy the tilting angle is restricted to a narrow range about $\frac{\pi}{2}$. In that sense, a high field does not lead to a more stable topological phase, although $E_\mathrm{Zee}>\sqrt{\Delta^2+\mu^2}$ is a necessary prerequisite\cite{Kitaev2001, LutchynSauDasSarma2010, OregRefaelOppen2010}. This is readily seen, since this latter condition acts on the energy gap at zero-momentum, which does not depend on the direction of the field. In contrast, if the phase transition is driven by $\vartheta$, the gap closes near the Fermi momentum\cite{OscaRuizSerra2014} at $\sqrt{\varkappa_1}$, cf. Fig.~\ref{FigEnergyLevels}, where increasing the field strength pushes the low-energy band closer to zero.

\section{Differential conductance characteristics}\label{SecConductance}
In the remainder of this paper, we focus on the differential conductance characteristics of a junction of the Majorana nanowire with a normal lead and the impact of tilting $\bm B$. To the best of our knowledge, the angular dependence of the differential conductance in such junctions has only been briefly discussed in Ref.~\onlinecite{LinSauDasSarma2012} so far, based on numerical studies of a tight-binding model. In contrast, we will analyze the current through the system in a simple continuum model. In the following, we will for simplicity set $\mu=0$.
\par We assume infinite wire length and a tunnel barrier of strength $V$ at the junction (located at $x=0$). The normal ($x<0$) and superconducting ($x>0$) sections of the wire are modeled with the same Hamiltonian Eq.~\eqref{EqHamiltonian}, where we just set $\Delta=0$ in the normal state. For electrons impinging from the normal side onto the junction we investigate the coefficients of reflected and transmitted waves. To solve the scattering problem, we employ a Blonder-Tinkham-Klapwijk (BTK) formalism\cite{BTK}, i.e., matching of wavefunctions at the junction. The original BTK scheme is extended to account for the spin as well.
\par At a given energy $E$, we first obtain all possible momenta by solving $p_k(E)=0$ for the normal and the superconducting wire. Exact diagonalization of Eq.~\eqref{EqHamiltonian} at each $k$ (including complex) then yields plane-wave states $\Psi_k(x) = \psi_k e^{ikx}$ with four-component spinors $\psi_k$. The incident electron wave $\Psi_{k_\mathrm{in}}^\mathrm{in}$ is always chosen from the normal low-energy band. All other states that correspond to incoming waves are discarded. The scattering process comprises ordinary and Andreev reflection into the normal lead, and transmission without ($k>0$) and with ($k<0$) branch crossing into the superconducting lead. The corresponding scattering coefficients are denoted $a_i,b_i,c_i,d_i$, respectively, where $i\in\{1,2\}$ labels the pseudospin. The total wavefunctions on the normal and superconducting side of the junction are then 
\begin{eqnarray}
\Psi^\mathrm{n}(x<0) &=& \Psi_{k_\mathrm{in}}^\mathrm{in}+\sum_{i=1,2}a_i\Psi_{k_{a,i}} + b_i\Psi_{k_{b,i}}, \\
\Psi^\mathrm{sc}(x>0) &=& \sum_{i=1,2}c_i\Psi_{k_{c,i}} + d_i\Psi_{k_{d,i}}.
\end{eqnarray}
At the junction, we impose the boundary conditions 
\begin{eqnarray}
\Psi^\mathrm{n}(x\rightarrow0^-) - \Psi^\mathrm{sc}(x\rightarrow 0^+) &=& 0\\ \partial_x\Psi^\mathrm{n}(x\rightarrow 0^-) - \partial_x\Psi^\mathrm{sc}(x\rightarrow 0^+) &=& \frac{2mV}{\hbar^2}\Psi(0),
\end{eqnarray}
and solve the resulting linear system of equations to obtain all scattering coefficients. The probability current 
\begin{equation}
  J = \frac{\hbar}{m}\mathfrak{Im}\left(\Psi^\dagger\partial_x\tau_z\Psi\right)
	   +\frac{\alpha}{\hbar}\Psi^\dagger\sigma_z\Psi
\end{equation}
carried by each outgoing wave, where we have taken into account a contribution due to the SOC\cite{Bottegoni2012}, is proportional to the square of the absolute value of the respective coefficient. Here, $\tau_z$ and $\sigma_z$ denote Pauli matrices acting in particle-hole and spin space, respectively. In the sub-gap regime, where the Majorana modes reside, the system is effectively spinless, therefore we will relinquish the distinction of states with different pseudospin for the discussion of the scattering probabilities, denoted $A, B, C,$ and $D$. Then, $C$, for instance, reads
\begin{equation}\label{EqCoefficient}
C = \sum_{i=1,2}|c_i|^2
    \frac{\left|\psi_{k_{c,i}}^\dagger\left(\mathfrak{Re}(k_{c,i})\tau_z
		   +\frac{\alpha m}{\hbar^2}\sigma_z\right)\psi_{k_{c,i}}\right|}
    {\left|k_\mathrm{in}+\frac{\alpha m}{\hbar^2}
		    \psi_\mathrm{in}^\dagger\sigma_z\psi_\mathrm{in}\right|}
.\end{equation}
Note that for $A$ and $B$ the term $\psi_k^\dagger\tau_z\psi_k$ gives always just $-1$ (holes, Andreev reflection) or $1$ (electrons, ordinary reflection), respectively. The differential conductance at $E$ through the junction at zero temperature is finally given by\cite{BTK,Prada2012} $\frac{dI}{dE} = 1+A-B$ in units of $\frac{e^2}{h}$, and inside the gap, where $C=D=0$, even simpler as $\frac{dI}{dE}=2A$ by conservation of probability ($A+B+C+D=1$).
\begin{figure}
\subfigure{\includegraphics[width=\columnwidth]{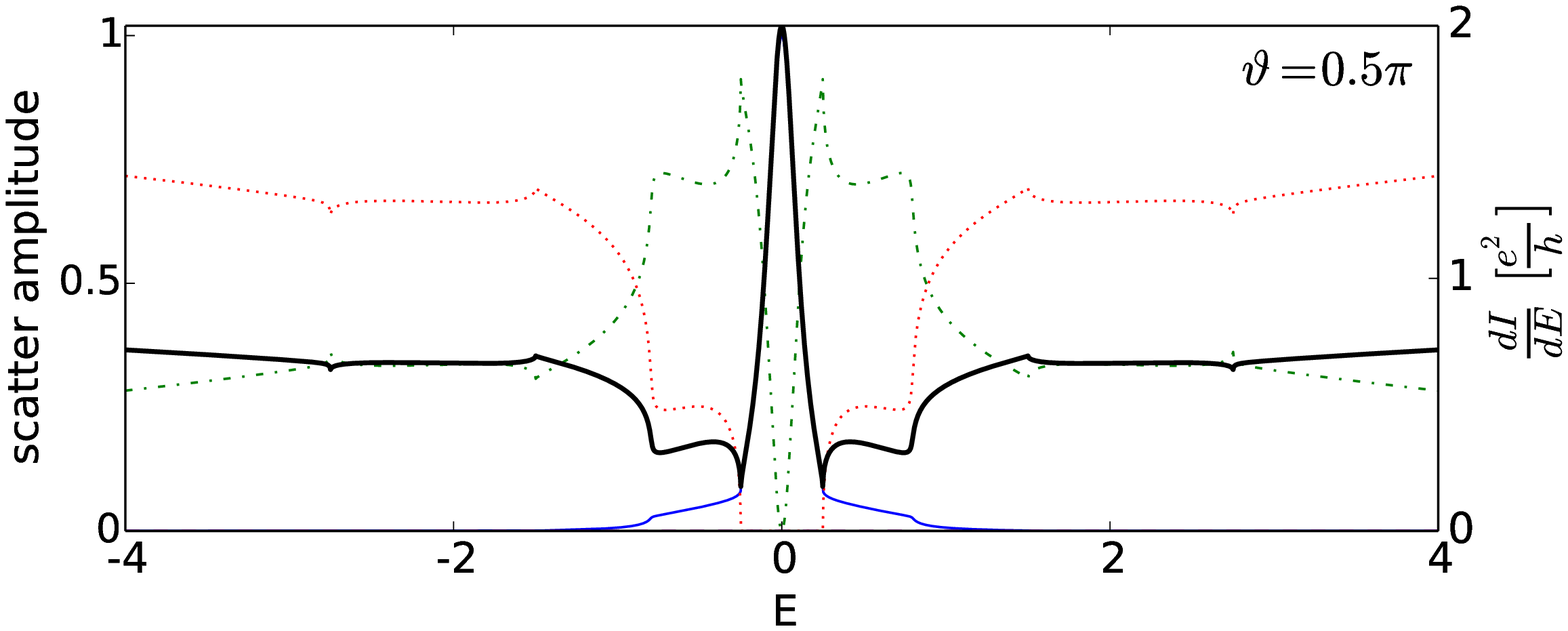}}\\
\subfigure{\includegraphics[width=\columnwidth]{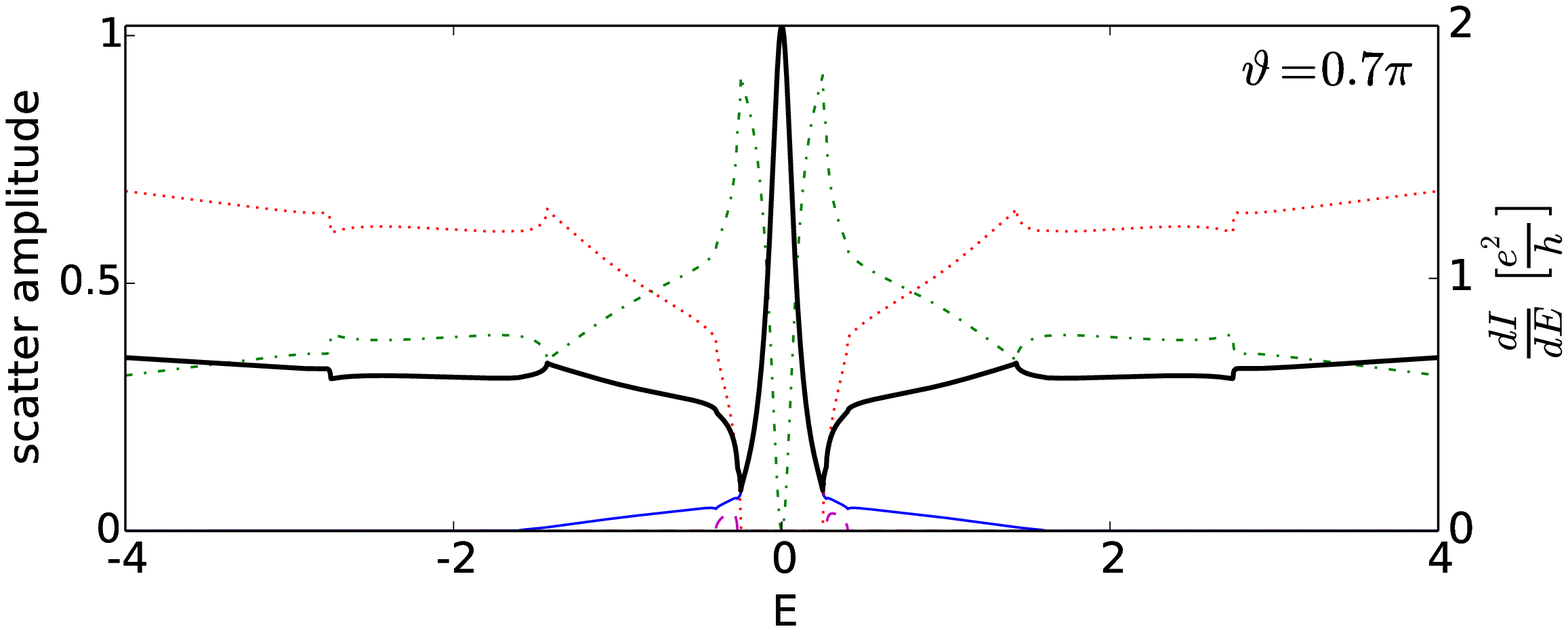}}\\
\subfigure{\includegraphics[width=\columnwidth]{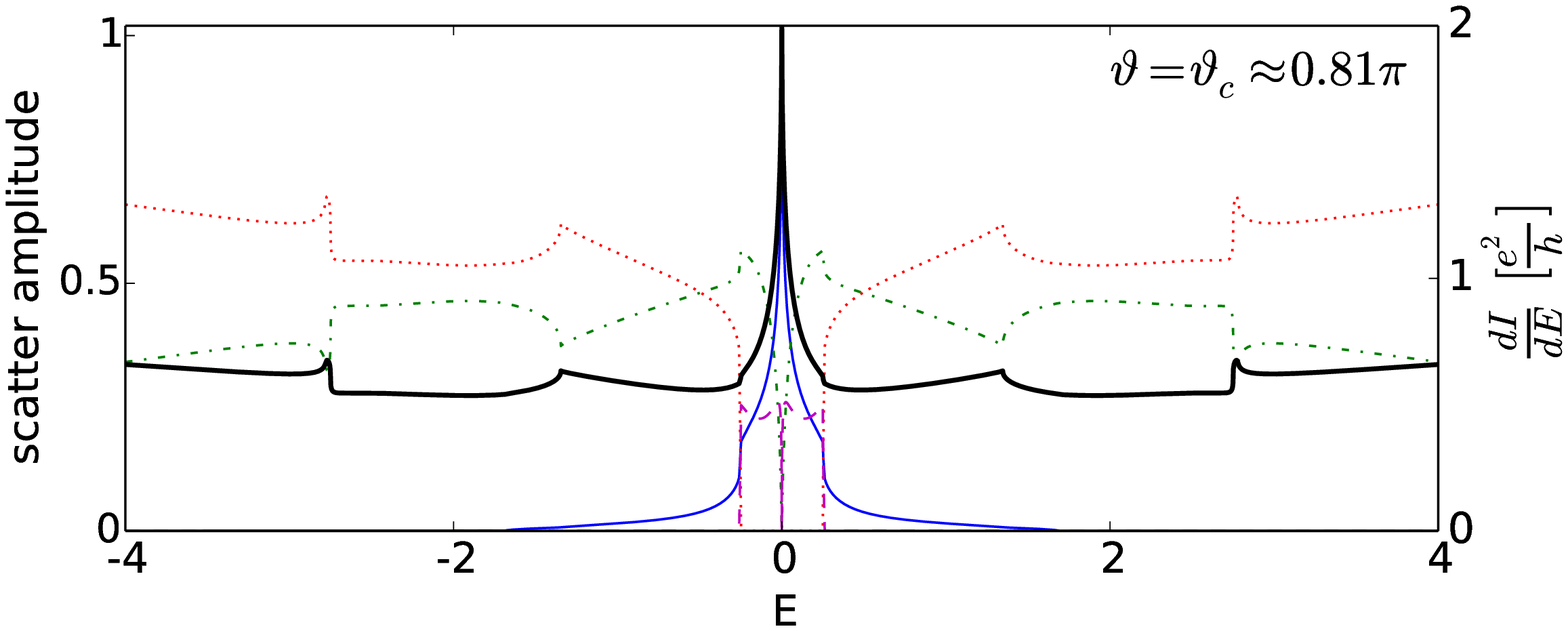}}\\
\subfigure{\includegraphics[width=\columnwidth]{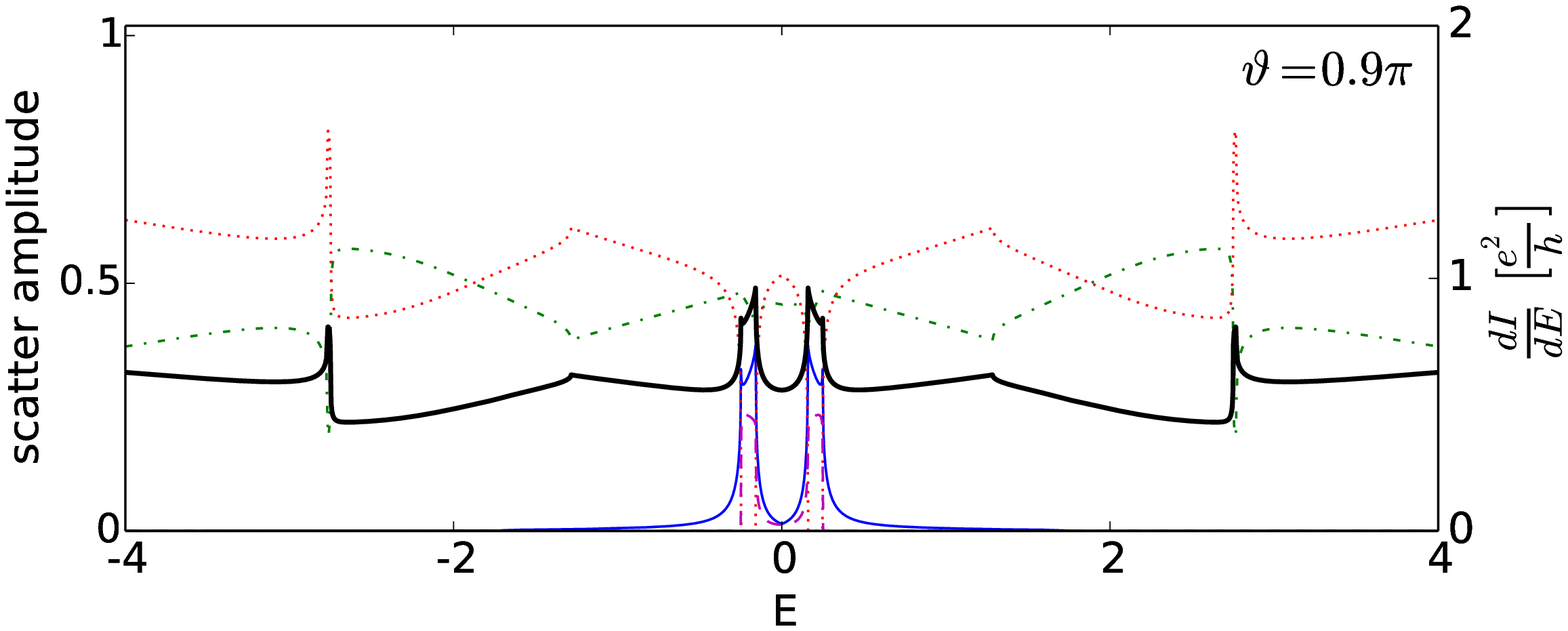}}\\
\subfigure{\includegraphics[width=\columnwidth]{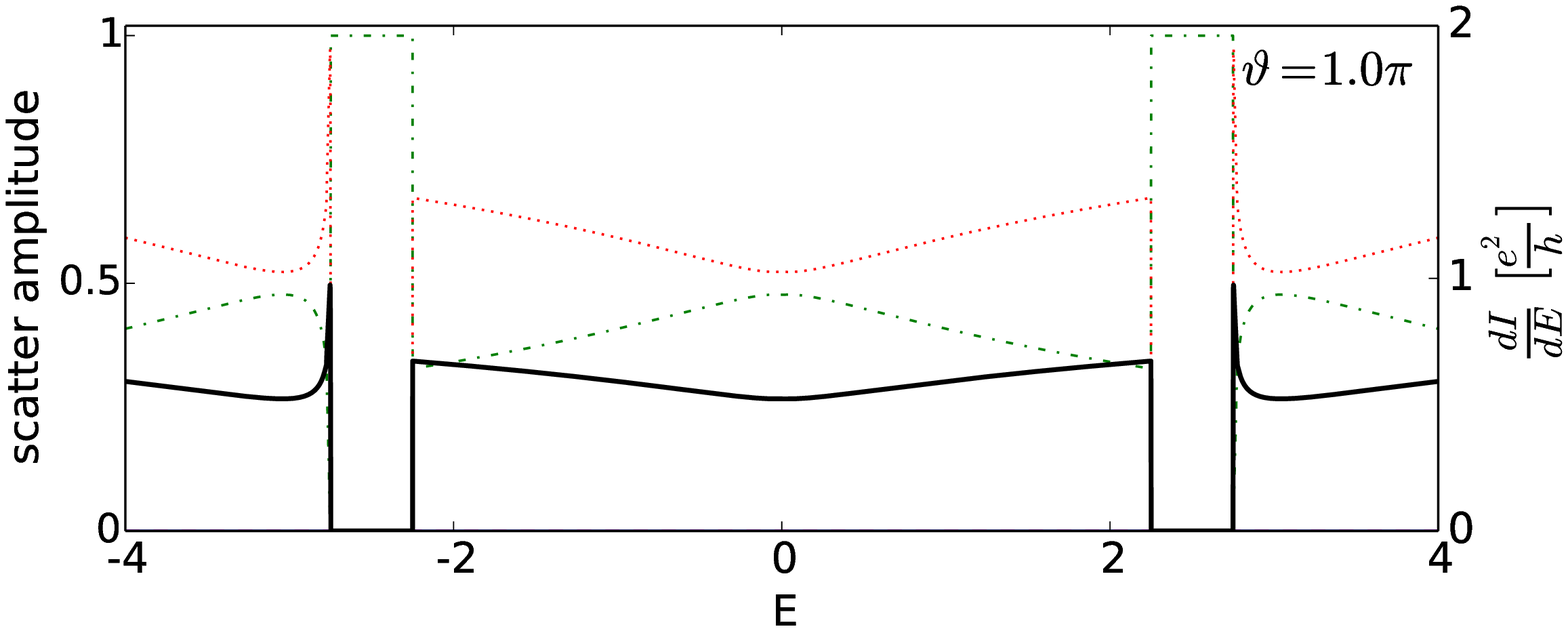}}
\caption{\label{FigDiffConductance}The energy-resolved scattering coefficients (left scale) and differential conductance characteristic (right scale) of a normal-Majorana nanowire junction at different tilting angles of the magnetic field: Andreev reflection $A$ (blue solid line), ordinary reflection $B$ (green dash-dotted line), transmission without branch crossing $C$ (red dotted line), transmission with branch crossing $D$ (purple dashed line), and differential conductance (black bold solid line). Parameters: $m=1, \Delta=1{.}25, E_\mathrm{Zee}=1{.}5, \alpha=\sqrt{1/2}, V=2{.}0$.}
\end{figure}
\par By this scheme, we obtain the scattering probabilities and the differential conductance profile $\frac{dI}{dV}(E)$ of the junction for different field directions, cf. Fig.~\ref{FigDiffConductance}. The scattering probability profiles indicate the features of the bandstructure at the respective respective angle, e.g. the gap width. In the topological phase, the conductance peak at zero energy that signals the existence of Majorana zero-modes is clearly seen. The peak gets narrower as the tilting angle of the field approaches the critical angle, and disappears in the trivial phase. As expected, the peak height exhibits the quantized value\cite{Sengupta2001, LawLeeNg2009, Flensberg2010, Wimmer2011} of $2\frac{e^2}{h}$ due to resonant Andreev reflection.
\par
\begin{figure}
\includegraphics[width=\columnwidth]{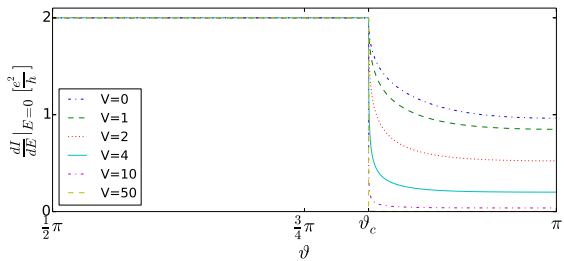}
\caption{\label{FigZBCP}The differential conductance at zero energy as a function of the tilting angle of the field for different tunnel barrier strengths $V$. Parameters: $m=1, \Delta=1{.}25, E_\mathrm{Zee}=1{.}5, \alpha=\sqrt{1/2}$.}
\end{figure}

 Attempts at detecting emergent Majorana zero-modes experimentally originally focused on the quantized value of the zero-energy differential conductance as the hallmark of such states. Under real conditions, however, only much smaller values are observed\cite{DelftExp, LinSauDasSarma2012}. Other, more qualitative and more robust distinguishing criteria are required. We propose that sharp change in the zero-energy differential conductance peak at the critical tilting angle $\vartheta_c$ of the field, provides an appropriate further qualitative criterion for examining the topological nature of measured signatures. In experiments, it may be difficult to record the full conductance profiles as in Fig.~\ref{FigDiffConductance} with the required precision. Therefore, we propose to measure the zero-energy differential conductance for different tilting angles of the field while varying the tunnel barrier strength of the junction. The predicted behavior is shown in Fig.~\ref{FigZBCP}. A qualitative change of the dependence of $\frac{dI}{dE}(0)$ on $V$ should be observed at the critical angle upon entering the trivial phase, where the conductance can be suppressed by increasing the tunnel barrier. In the topological state, the value of the zero-bias conductance peak is impervious to the change in barrier strength, being protected by topology. 
\par At finite temperatures well below the superconducting transition temperature, the impact on the results in Fig. \ref{FigDiffConductance} is to slightly smear the sharp cusp at $\vartheta_c$. The main change in qualitative behavior above and below $\vartheta_c$ is robust. The main effect on the critical angle itself can be accounted for by taking into account the temperature dependence of the gap in Eq. \ref{Crit_Ang}. Finite-size effects are also present, in principle. A finite length of the Majorana nanowire causes an overlap of the exponentially localized topological states at the ends of the wire\cite{LinSauDasSarma2012}. Thus, the transition happens before the low-energy band reaches zero and the true topological regime is expected to be slightly narrower than predicted by $\vartheta_c$. Numerical data from Ref.~\onlinecite{OscaRuizSerra2014} indicate, however, that this effect is not important.

\section{Conclusion}\label{SecConclusion}
In this paper, we have studied semiconductor nanowires with SOC and $s$-wave superconductivity in an external magnetic field with arbitrary direction in an analytically accessible continuum model. We have derived the critical tilting angle $\vartheta_c$ of the field relative to the SOC direction, at which the topological (Majorana) zero-modes disappear. Our result confirms recent numerical findings\cite{OscaRuizSerra2014}. Furthermore, we have considered normal-Majorana nanowire junctions and obtained the differential conductance characteristics at various angles, where, as expected, a stable peak at zero-energy with the quantized value of $2\frac{e^2}{h}$ occurs as long as the field is not tilted beyond the critical angle $\vartheta_c$. The peak disappears for fields aligned too much in the direction of the SOC and the value of the zero-energy differential conductance becomes strongly dependent on the tunnel barrier strength. We have pointed out the qualitative change of the dependence on the barrier strength at the critical angle and suggest it as further criterion to test the topological nature of the experimentally observable signals, even if the theoretical quantized peak value may not be reached under realistic conditions.

A.S. and S.R. acknowledge support from the Norwegian Research Council, Grants 205591/V20 and 216700/F20. We thank Jacob Linder for helpful comments.  
\bibliography{references}

\end{document}